\documentclass[twocolumn]{aa}

  \usepackage{amscd}
  \usepackage{amsmath}
  \usepackage{amssymb}
  \usepackage[T1]{fontenc}
  \usepackage[varg]{txfonts}
  \usepackage{graphicx}
  \usepackage{natbib}
  \usepackage{verbatim}
  \usepackage{array}

  \newcommand{\mbf}[1]{\ensuremath\mbox{\boldmath{$#1$}}}
  \newcommand{\dd}{\ensuremath \mathrm{d}}
  \newcommand{\DD}{\ensuremath \mathrm{D}}
\begin{document}

\title{Following multi-dimensional Type Ia supernova explosion models
  to homologous expansion}
\titlerunning{Following multi-D SN Ia explosion models to homologous expansion}

\author{F. K. R{\"o}pke}
   \institute{Max-Planck-Institut f\"ur Astrophysik,
              Karl-Schwarzschild-Str. 1, D-85741 Garching, Germany\\
              \email{fritz@mpa-garching.mpg.de}
             }

\abstract{The last years have witnessed a rapid development of
  three-dimensional models of Type Ia supernova
  explosions. Consequently, the next step is
  to evaluate these models under variation of the initial parameters
  and to compare them with observations. To calculate synthetic lightcurves
  and spectra from numerical models, it is mandatory to follow the
  evolution up to homologous expansion. We report on methods to
  achieve this in our current implementation of multi-dimensional Type
  Ia supernova explosion models. The novel scheme is thoroughly tested
  in two dimensions and a simple example of a three-dimensional
  simulation is presented. We discuss to what degree the
  assumption of homologous expansion is justified in these models.

\keywords Stars: supernovae: general -- Hydrodynamics -- Methods: numerical}

\maketitle


\section{Introduction}
\label{intro_sect}

Type Ia supernova (SN Ia) explosions are commonly attributed to
thermonuclear explosions of white dwarf (WD) stars \citep{Hoyle1960a}
in a binary system. The precise scenario of these events is, however,
controversial \citep[for a recent review
see][]{hillebrandt2000a} and it seems well possible that different
mechanisms contribute to the SN Ia class. The final decision must be
made by comparing theoretical models with detailed
observations. Substantial progress has been achieved on both sides during the
past years. The rapid development of three-dimensional models of
thermonuclear supernova explosions \citep[e.g.][]{hillebrandt2000b,
reinecke2002b, reinecke2002d, reinecke2002c, gamezo2003a, calder2004a}
leads naturally to the question whether they are capable of
reproducing observed features of SNe Ia. Systematic parameter studies
on the basis of three-dimensional SN Ia explosion models have become
possible \citep{roepke2004c}.

For the models by \citet{reinecke2002b, reinecke2002d, reinecke2002c}
synthetic light curves have been calculated with very
encouraging results \citep{sorokina2003a}. However, for a thorough
comparison of SN Ia models with observation via synthetic light
curves and spectra, it is inevitable to follow the explosion
simulations up to homologous expansion. The models by
\citet{reinecke2002b, reinecke2002d, reinecke2002c} reach only to
$t=1.5 \, \mathrm{s}$ and also
\citet{gamezo2003a} stayed below $t=2 \, \mathrm{s}$, although it seems
likely that the former are more advanced in the explosion stage,
since there the initial flame shape is superposed by stronger spatial
perturbations which enhance the development of Rayleigh-Taylor like
instabilities and turbulence.

A further motivation for
evolving the explosion models to later times is to explore the effects
of slower turbulent combustion regimes (the so-called distributed
burning). These apply to low fuel densities reached after $t \sim
1.5 \, \mathrm{s}$ by expansion of the WD. The present work focuses on
reaching the stage of homologous expansion and the issue of late
burning will be addressed in a separate study.

In the following we report on modifications of the code of
\citet{reinecke1999a, reinecke2002b} that enable us to follow the
explosion model for much longer times than previous multidimensional
simulations. These
novel methods are tested in two-dimensional models. For these as
well as for an example in 
three dimensions the approach to homologous expansion is studied.
Our simulations are based on the
\emph{single degenerate Chandrasekhar mass deflagration scenario}
\citep[see][]{hillebrandt2000a}. However, the methods developed here can
easily be applied to other models, featuring for instance a delayed
detonation.

\section{Homologous expansion in SN Ia explosions}

A few seconds after ignition of the flame SN Ia explosions are expected
to approach
self-similar (homologous) expansion, which is characterized by a fluid
velocity proportional to the radius,
\begin{equation} \label{homolog_eq}
v(r) \propto r.
\end{equation}
This implies that the relative positions of the fluid elements do not
change anymore. Consider the equation of motion in Lagrangian
formulation,
\begin{equation} \label{eom_lagr_eq}
\rho \frac{\mathrm{d}\mbf{v}}{\mathrm{d} t} = - \mbf{\nabla}P - \rho
\mbf{\nabla} \Phi,
\end{equation}
with $P$, $\rho$, and $\Phi$ denoting the pressure, the density and
the gravitational potential, respectively. From this equation it
follows that no relative change of the
velocity of the fluid elements co-moving with the expansion is
equivalent to a vanishing pressure gradient and evanescent
gravitational force. Hydrodynamical interaction ceases if $\mbf{\nabla}P = 0$ and
free expansion is reached for $\mbf{\nabla}\Phi \rightarrow 0$.

Another criterion for the approach to homologous expansion can be
derived from the energy balance of the explosion process. All energy
liberated in the explosion is generated by thermonuclear
reactions. While only a tiny fraction of this energy is radiated away
giving rise to the observable luminous event, most of it is used to
overcome the gravitational
binding of the WD star and to accelerate the ejecta. Due to expansion
the ejecta are diluted and cooled. Thus, approaching homologous
expansion, the gravitational energy and the internal energy should
become small compared to the kinetic energy.

\section{Numerical model and modifications}

\subsection{Explosion model}

Our numerical model is based on the scheme proposed by
\citet{reinecke1999a} and includes the improvements described by
\citet{reinecke2002b}. 

The hydrodynamics is modeled by the Euler
equations with species conversion and appropriate source terms to take
into account nuclear reactions. 
The numerical solution of these equations is based on the
\textsc{Prometheus} implementation \citep{fryxell1989a} of the
Piecewise Parabolic Method (PPM) by \citet{colella1984a}. 

The equation of state (EoS) models the degenerate matter of the WD star
taking into account an electron gas that is degenerate and
relativistic to a variable degree, the completely ionized nuclei
following the Maxwell-Boltzmann distribution, a photon gas and
electron-positron pair creation.

The appropriate description of the nuclear reactions would be given by a full
reaction network. Since this is too costly to run concurrently with the
explosion model, a simplified description is chosen.  Following the
approach suggested by \citet{reinecke2002b} five species are included,
namely $\alpha$-particles, $^{12}$C, $^{16}$O, $^{24}$Mg as a
representative of intermediate mass elements, and $^{56}$Ni as a representative
of iron group nuclei. The composition of the WD prior to the explosion is
assumed to be a mixture of $^{12}$C and $^{16}$O. At the
initially high densities burning proceeds to nuclear statistical equilibrium
(NSE) composed of $\alpha$-particles and
nickel. Depending on temperature and density in the ashes, the
fraction of $\alpha$-particles and nickel changes. Once the
fuel density drops below
$5.25 \times 10^7 \,\mathrm{g}\,\mathrm{cm}^{-3}$ due to the expansion
of the WD, burning is assumed
to terminate at intermediate mass elements and below $1 \times 10^7
\,\mathrm{g}\,\mathrm{cm}^{-3}$ burning becomes very slow and is not
followed anymore. This may not be a satisfactory approach when the
evolution is simulated to later times since it is possible that even
the much slower flame contributes to the production of intermediate
mass elements. However, this will not affect the production of iron group
elements and may therefore be neglected in order to obtain a first-order
estimate of the produced nickel mass
and the total energy. But for synthetic spectra it may have a
significant impact. We will address this issue in a forthcoming study and
ignore it in the following, focusing on the hydrodynamical
evolution of the models rather than on a detailed nucleosynthetic
description.

Since the range of relevant length scales in a SN
Ia explosion (from the radius of the WD of $\sim$$1000 \, \mathrm{km}$
to the width of the flame which is less than a centimeter) is by far
too large to be resolved in numerial 
simulations, the flame evolution has to be modeled by an appropriate
effective scheme. 
This is accomplished by treating the flame as a discontinuity between fuel and
ashes. We will call this model representation of the flame ``flame
front'' in the following. 

In the so-called \emph{flamelet regime} of turbulent combustion, which
applies to large parts of thermonuclear SN explosions \citep{niemeyer1997b}, the flame is
wrinkled by interaction with turbulent velocity fluctuations. These
stem from a turbulent cascade, which is generated by large scale
buoyancy (Rayleigh-Taylor) instabilities. The interaction with turbulence
increases the surface of the flame front and accelerates its
propagation. Turbulence dominates flame propagation down to the Gibson
scale, which is far below the numerical resolution. Below the Gibson
scale flame propagation proceeds stably in the \emph{cellular burning
  regime} and therefore these scales have little effect on the
macroscopic flame evolution
\citep{roepke2003a,roepke2004a,roepke2004b}.

According to these considerations the discontinuity description not only
averages the finite
internal flame structrue but additionally defines a mean flame
location neglecting the wrinkling of the flame at scales below the
resolution of the computational grid. In this way it represents a
``flame brush'' rather than a smooth flame.
To determine the turbulent burning velocity of the flame front the
sub-grid scale model described by \citet{niemeyer1995b} is applied.

The propagation of the flame front is implemented following the  the
level set technique by \citet{osher1988a}.  This technique was extended to
deflagration flames by \citet{smiljanovski1997a} and a scheme (the
so-called ``passive implementation'') applicable to SN Ia explosion
models was developed by \citet{reinecke1999a}. The basic idea is to
associate the flame front $\Gamma$ with the zero level set of a signed distance
function $G$ 
$$
\Gamma := \{ \mbf{r} \,|\, G(\mbf{r}) = 0\}, \qquad |\mbf{\nabla}G| \equiv 1.
$$
Its temporal evolution is prescribed by the so-called $G$-equation
$$
\frac{\partial G}{\partial t} = - \mbf{D}_\mathrm{f} \cdot \mbf{\nabla}G.
$$
The velocity of the front motion is given by 
\begin{equation}\label{flame_vel_eq}
\mbf{D}_\mathrm{f} =
\mbf{v} + s_\mathrm{t}\mbf{n}
\end{equation}
for the passive implementation with
$\mbf{v}$ and $\mbf{n}$ denoting the fluid velocity and normal vector
to the flame front, respectively. The turbulent
burning velocity  $s_\mathrm{t} = \sqrt{2 q_\mathrm{sgs}}$ is
determined by the subgrid-scale energy $q_\mathrm{sgs}$ provided by
the sub-grid scale turbulence model.
With this approach the flame front is described as a sharp
discontinuity and topological changes are handled without problems
\citep{reinecke1999a}. 

\subsection{Co-expanding computational grid}

To evolve numerical models to homologous expansion it is necessary to
capture significantly growing length scales in the computational
domain. One approach could be to start with a wide-spaced static
computational grid and use adaptive mesh refinement to resolve the
relevant parts of the domain. This approach will, however, become
increasingly expensive with evolution time, since the developing
turbulence will fill large parts of the domain in which the grid has
to be refined. In explicit schemes the maximum time evolution for each
hydro step is
given by the CFL criterion \citep{courant1928a}. It states that to
ensure stability of the hydrodynamics solver the time step needs to be
chosen smaller than the time the fastest possible wave would need to
cross a computational grid cell. Thus refining the grid drastically
increases the number of hydro steps necessary to evolve the model.
The level set method together with the sub-grid scale model provide
a high accuracy even at relatively coarse computational grids.
Therefore we do not apply adaptive mesh refinement. 

The simulations by \citet{reinecke2002b,
reinecke2002d, reinecke2002c} were performed on a grid that was equally
spaced in the center of the WD. To capture at least parts of the
expansion of the WD, the width of the computational cells was
exponentially increased in 
the outer regions resulting in a highly non-uniform grid. In the
following we will refer to this implementation as
the \emph{static grid approach}. An advantage of this approach is that
the resolution of the flame front is relatively fine as long as it
stays in the uniform inner part of the grid. If this is true for the
stage of the highest energy generation rate, numerical convergence of
the models can be reached already for a grid with 256 cells per spatial
dimension \citep{reinecke2002c}. The static grid approach, however,
has a number of disadvantages:
\begin{itemize}
\item With the static grid it is very expensive to follow the
  explosion model to homologous expansion.
\item The non-uniformity of the grid is an obstacle to the
  implementation of the level set method. The necessary re-initialization of the
  signed distance function $G$ \citep{sussman1994a} is difficult in
  the elongated grid cells in the outer regions.
\item Some more advanced sub-grid scale models
  tested in simulations similar to that presented below (Schmidt et
  al., in preparation) require a uniform computational grid.
\item In the enlarged outer computational cells the flame front
  representation is very 
  coarse. This problem is less severe in models that are confined to
  only one spatial octant (which was the case in all previous
  simulations), because the buoyancy-induced burning bubbles developed
  preferentially along the axes. This is not true if this artificial
  symmetry constraint is abolished and simulations of the full WD
  star are performed \citep{roepke2004d}.
\end{itemize}

Therefore we implemented a computational grid that co-expands with the
explosion. Here the grid spacing is allowed to vary and the resulting
discretization of the equations of hydrodynamics is neither Eulerian nor
Lagrangian. With this scheme it
is possible to track the expansion of the WD during the SN Ia
explosion with a uniform computational grid. This approach will be
denoted as \emph{co-expanding grid} in the following. It overcomes the
drawbacks of the static grid approach. 

However, due to the uniformity and the expansion of the grid, the flame front will
be less resolved at the stage of maximum energy generation if one
starts out with the same grid spacing as in the
inner parts of the static grid approach.
Therefore it has to be tested
which resolution is necessary in the new implementation to reach
numerical convergence.

Our approach is based on the ``moving grid''
technique\footnote{Originally, \citet{winkler1984a} called this
technique \emph{adaptive mesh}. To avoid confusion with adaptive
mesh \emph{refinement} methods, we will denote it \emph{moving grid
technique}, following \cite{mueller1994a}.} by
\citet{winkler1984a}. This technique was already implemented in the
original \textsc{Prometheus} code \citep{fryxell1989a}. We will give a
brief outline of the derivation
of the underlying equations following \citet{mueller1994a} and
\citet{winkler1984a}. Let $(\partial / \partial t)$ denote the Eulerian
derivative taken with respect to fixed coordinates in the laboratory
frame and $(\DD / \DD t)$ the Lagrangian derivative taken with respect
to a definite fluid element. The derivative with respect to fixed
values of the moving grid (``moving grid derivative'') will be
designated $(\dd / \dd t)$. Analogous to the fluid velocity
$$
\mbf{v} = \frac{\DD}{ \DD t} \mbf{r}_\mathrm{e},
$$
where $\mbf{r}_\mathrm{e} = \mbf{r}_\mathrm{e}(\mbf{r}_0, t)$
specifies the position of a definite fluid element, we define the
\emph{grid velocity}
$$
\mbf{v}_\mathrm{grid} = \frac{\dd}{\dd t} \mbf{r}_\mathrm{grid}.
$$
Here, $\mbf{r}_\mathrm{grid}$ is the position of a definite set of grid
coordinates. The relative velocity between fluid and moving grid is
then given by
$$
\mbf{v}_\mathrm{rel} = \mbf{v} - \mbf{v}_\mathrm{grid}.
$$
Lagrangian and Eulerian coordinates are special cases of the moving
grid for $\mbf{v}_\mathrm{grid} = \mbf{v}$ and $\mbf{v}_\mathrm{grid}
= 0$, respectively. The Eulerian and Lagrangian time derivatives of
the density $a$ of an extensive quantity are related by
$$
\frac{\DD}{\DD t} a = \frac{\partial}{\partial t} a + \mbf{v}
\mbf{\nabla} a,
$$
with $\mbf{\nabla}$ taken with respect to Eulerian
coordinates. Similarly,
$$
\frac{\dd}{\dd t} a = \frac{\partial}{\partial t} a + \mbf{v}_\mathrm{grid}
\mbf{\nabla} a
$$
relates the moving grid derivative to the Eulerian derivative. 

Introducing the Jacobian $J_\mathrm{f}$ of the transformation between
coordinates defining the initial volume of a fluid element $\dd
V^0_\mathrm{fluid}$ and the volume $\dd V_\mathrm{fluid} =
J_\mathrm{f} \dd V^0_\mathrm{fluid}$ of same fluid element
at later times it is straightforward to derive the Euler expansion
formula (e.g.\ \citealt{mueller1994a})
$$
\frac{\DD \ln J_\mathrm{f}}{\DD t} = \mbf{\nabla} \cdot \mbf{v},
$$
which leads to the Reynolds transport theorem
$$
\frac{\DD}{\DD t} \int_{V_\mathrm{fluid}} \left( a \, \dd
  V_\mathrm{fluid}\right) =
\int_{V_\mathrm{fluid}}\left\{ \frac{\partial a}{\partial t} +
  \mbf{\nabla} \cdot (\mbf{v} a) \right\} \dd V_\mathrm{fluid}.
$$
Analogously one obtains
the \emph{moving grid expansion formula}
$$
\frac{\dd \ln J}{\dd t} = \mbf{\nabla} \cdot \mbf{v}_\mathrm{grid},
$$
where $J$ now denotes the Jacobian of the transformation between a
moving grid volume  $\dd V_\mathrm{grid} =
J_\mathrm{f} \dd V^0_\mathrm{grid}$ and its initial volume $\dd
V^0_\mathrm{grid}$. 

This leads to the \emph{moving grid transport
  theorem} (see e.g.\ \citealt{mueller1994a})
\begin{equation}\label{mgtt_eq}
\frac{\dd}{\dd t} \int_{V_\mathrm{grid}} \left( a \, \dd
  V_\mathrm{grid} \right) =
\int_{V_\mathrm{grid}}\left\{ \frac{\partial a}{\partial t} +
  \mbf{\nabla} \cdot (a \mbf{v} ) \right\} \dd V_\mathrm{grid}.
\end{equation}

Inserting the general form of a balance
equation in Eulerian form
$$
\frac{\partial}{\partial t} a = - \mbf{\nabla}\cdot (a\mbf{v}) + s(a),
$$
where $s(a)$ denotes the source term, into Eq.~(\ref{mgtt_eq}) one
obtains
\begin{equation*}
\begin{split}
\int_{V_\mathrm{grid}} s(a) \, &\dd V_\mathrm{grid}\\&=
\frac{\dd}{\dd t} \int_{V_\mathrm{grid}} a \, \dd
  V_\mathrm{grid} +
\int_{V_\mathrm{grid}} \mbf{\nabla} \cdot \left[ a (\mbf{v} -
\mbf{v}_\mathrm{grid}) \right] \dd V_\mathrm{grid}\\
&=\frac{\dd}{\dd t} \int_{V_\mathrm{grid}} a \, \dd
V_\mathrm{grid} + \iint_{\partial V_\mathrm{grid}} a
\mbf{v}_\mathrm{rel}\cdot\mbf{\dd S}.
\end{split}
\end{equation*}
With the appropriate quantities $a$ and source terms $s(a)$ this
defines the set of equations of hydrodynamics that is solved
numerically. 

\section{Simulation setup}

The two-dimensional simulations we report on, were carried out on one
quadrant assuming axial symmetry and mirror symmetry to the other
half space. 
In the three-dimensional simulations, only one spatial octant was
calculated assuming mirror symmetry to the other octants. All
simulations with co-expanding grid started with a computational domain
of $[2.02 \times 10^8 \, \mathrm{cm}]^2$ discretized on the respective
number of computational cells. Reflecting boundary conditions were
applied on all sides of the domain.

The explosion simulation was started with a WD near the Chandrasekhar
mass limit in hydrostatic equilibrium. Its initial temperature was
set to $T_0 = 5 \times 10^5 \, \mathrm{K}$. For the composition
a 50\% mixture of $^{12}$C and $^{16}$O was chosen. The
central density $\rho_c$ was set to $2.9 \times 10^9 \, \mathrm{g}
\mathrm{cm}^{-3}$. With these initial values a WD model was constructed
by integrating the equations of hydrostatic equilibrium using the
EoS described above (see
\citealt{reinecke_phd}). An important
point is that due to the limited range of the EoS this integration was
stopped as soon as a density of $10^{-3} \, \mathrm{g}
\mathrm{cm}^{-3}$ was reached.  This value was kept constant in the
outer regions (``pseudo-vacuum''), which thus cannot maintain hydrostatic equilibrium. The
resulting fluid motions in these parts of the
computational domain leave the physical evolution of the
explosion unaffected, since here the masses are very small. As will be discussed
below, they can, however, lead to numerical difficulties.
A WD of the mass of $2.797 \times 10^{33}
\, \mathrm{g}$ ($1.406 \, M_\odot$) resulted from the setup procedure.

The flame was ignited near the center of the WD in a sphere which was
superimposed by toroidal perturbations to accelerate the growth of
Rayleigh-Taylor like instabilities. This corresponds to the
\textit{c3} model of \citet{reinecke2002c}. This initial condition
is rather artificial. However, the exact way in which the flame
formation proceeds is not yet precisely known (\citet{garcia1995a} and
\citet{woosley2004a} provide
approaches), and it seems likely that it cannot be resolved in the kind
of simulations presented in the following.
Therefore one has to find an averaged initial flame
prescription. For the exemplary cases, our simple model will suffice, but
the initial condition of the burning front remains an open question of SN
Ia explosion models.

\begin{figure*}[p]
\centerline{
\includegraphics[height = 0.95 \textheight]
  {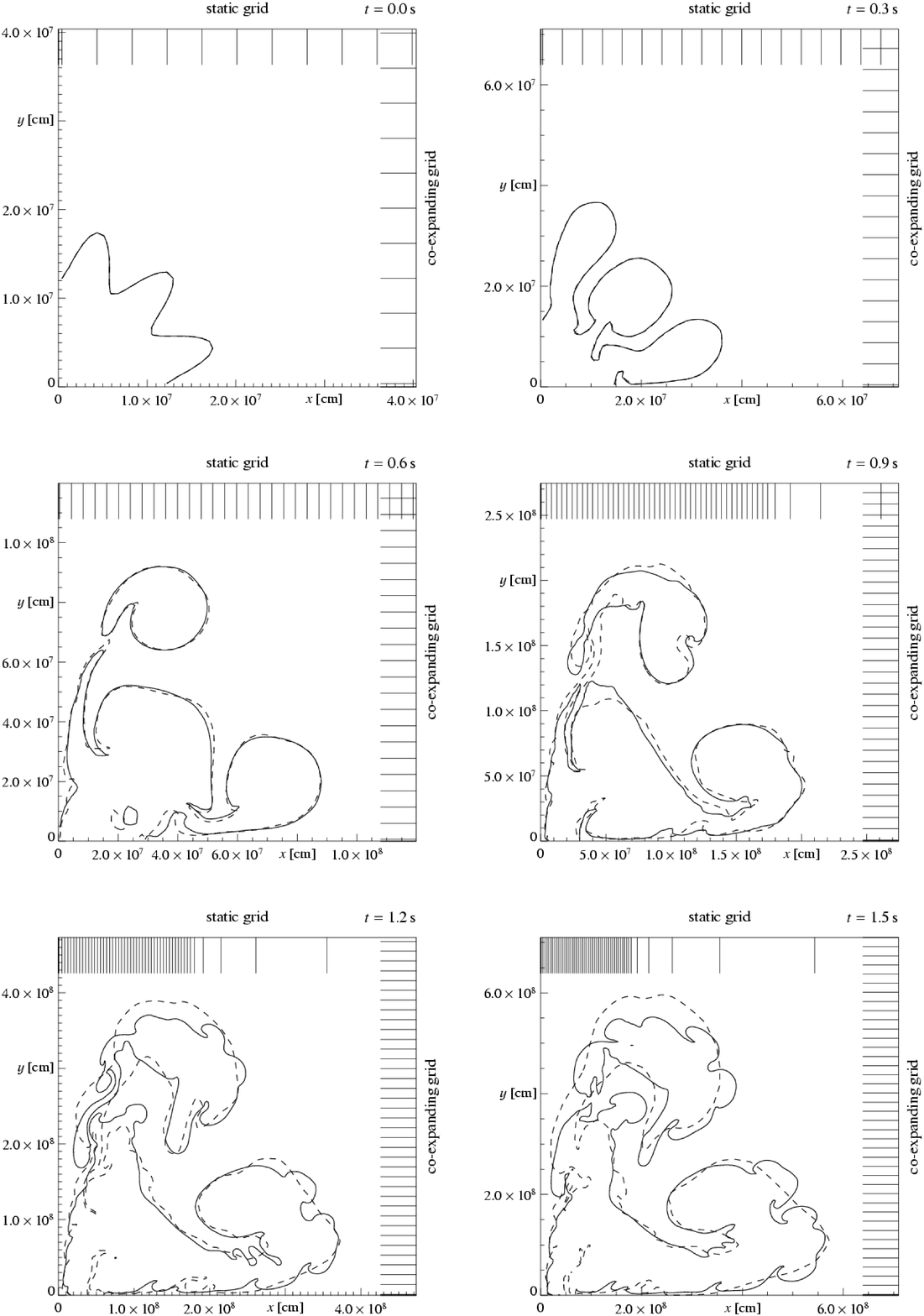}
}
\caption{Comparison of the flame front evolution in a simulation with
  co-expanding computational grid (solid) with a static grid simulation
  (dashed). The right and the upper axis ticks indicate the location of
  every fifth grid cell for the co-expanding and the static grid,
  respectively. \label{compare_fig}}
\end{figure*}

An issue that deserves some consideration is the prescription of the
grid expansion velocity. The goal is to follow the evolution of the WD
and thus the radial velocity of the edge of the star has to be
tracked. However, in our setup as described above, it is problematic to
define the edge of the WD. Neither the pressure nor the temperature
vanish here. A certain low density is also not a good indicator
of the WD's edge since the overall density drops with expansion and
especially the low density of the ashes can become comparable to that
``outside'' of the star. We therefore followed the $1.4 \, M_\odot$
mass shell. This was achieved by defining a belt that includes $0.006
\, M_\odot$ around this mass shell and determining the mean fluid
velocity inside this belt. According to this mean velocity the grid
coordinates were expanded self-similarly so that the $1.4 \, M_\odot$
mass shell was located at a fixed position at our computational
grid. By tracking a very high mass shell one can prevent
the star from reaching the domain boundaries even in case of strong
asphericities.

\section{Two-dimensional simulations}
\subsection{Comparison with static grid calculations}

In Fig.~\ref{compare_fig} the flame front morphologies in a co-expanding
grid calculation are compared to a static grid simulation at different
epochs of the explosion process. Both simulations start out from the
same flame shape and were carried out on an $[256]^2$ cell
domain. Until $t = 0.3 \, \mathrm{s}$ the evolution of both
flame fronts is almost identical. This is trivially the case because
the grid spacing is the same in the part of the domain that is occupied by the
flame front. In the co-expanding grid simulation the grid
spacing is changed according to the velocities around the mass shell
of $1.4 \, M_\odot$. This shell is reached by the first shock wave
due to the prompt incineration of the material behind the initial flame
after $t\approx 0.3 \, \mathrm{s}$. From that moment on the grid keeps
expanding self-similarly. Therefore its spacing at $t=0.6 \,
\mathrm{s}$ is larger than that of the inner part of the static
grid. The better resolution of the flame front in the static grid case leads
to a slight deviation of the flame front shape evolution (cf.\
Fig.~\ref{compare_fig}). However, at $t=0.9 \, \mathrm{s}$ the flame front
in the static grid simulation has already entered the part of the
domain with exponentially increasing grid spacing and consequently its
outer parts are less resolved than the flame front in the co-expanding grid
simulation. Therefore the flame front in the latter case develops more
features in the outer parts (see the plots corresponding to $t=1.2\,
\mathrm{s}$ and $t=1.5\, \mathrm{s}$ in Fig.~\ref{compare_fig}). 

Summarizing the results plotted in Fig.~\ref{compare_fig}, we find that
the flame front in the co-expanding grid simulation develops less
structure in the inner parts and more features in the outer parts than
the corresponding flame front in the static grid simulation. However, the
overall flame front evolution is very similar. 

\begin{figure}[t]
\centerline{
\includegraphics[width = \linewidth]
  {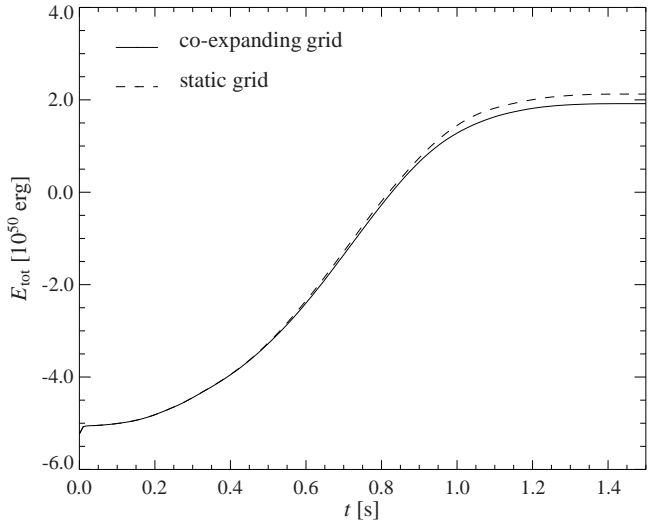}
}
\caption{Total energy in the simulations with co-expanding and static
  grid. \label{etot_compare_fig}}
\end{figure}

\begin{figure}[t]
\centerline{
\includegraphics[width = \linewidth]
  {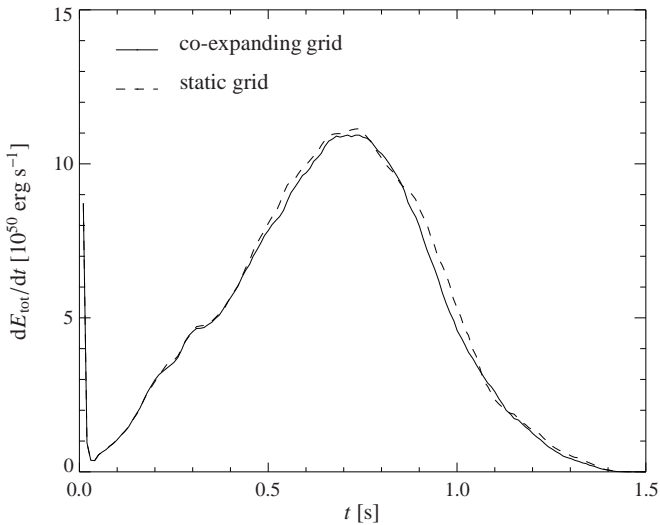}
}
\caption{Energy production rate in the simulations with co-expanding
  and static grid. \label{egen_compare_fig}}
\end{figure}

The total energies produced by the two models are plotted in
Fig.~\ref{etot_compare_fig}. Both models produced too little energy to
account for a SN Ia explosion. This is expected for two-dimensional
simulations with the chosen flame setup. The co-expanding grid
simulation produced slightly less energy than the static grid case. This
effect can again be attributed to the different evolution of the
resolution of the flame front. Figure~\ref{egen_compare_fig} shows clearly
that the energy production rate reaches a maximum at around $0.7 \,
\mathrm{s}$ (The initial high value of the energy generation
originates from the instantaneous incineration of the material behind
the initial flame.). At this time, however, the grid spacing at the
flame front
location in the co-expanding case is already wider than the one in the
static grid approach.

From our comparison we conclude that the
co-expanding grid scheme provides a reliable and robust model of the SN
Ia deflagration phase. Small differences compared to the previously
used static grid approach can be attributed to the different resolutions of the
flame front at different epochs of the explosion process.

\subsection{Numerical convergence}

\begin{figure*}[p!]
\centerline{
\includegraphics[width = \linewidth]
  {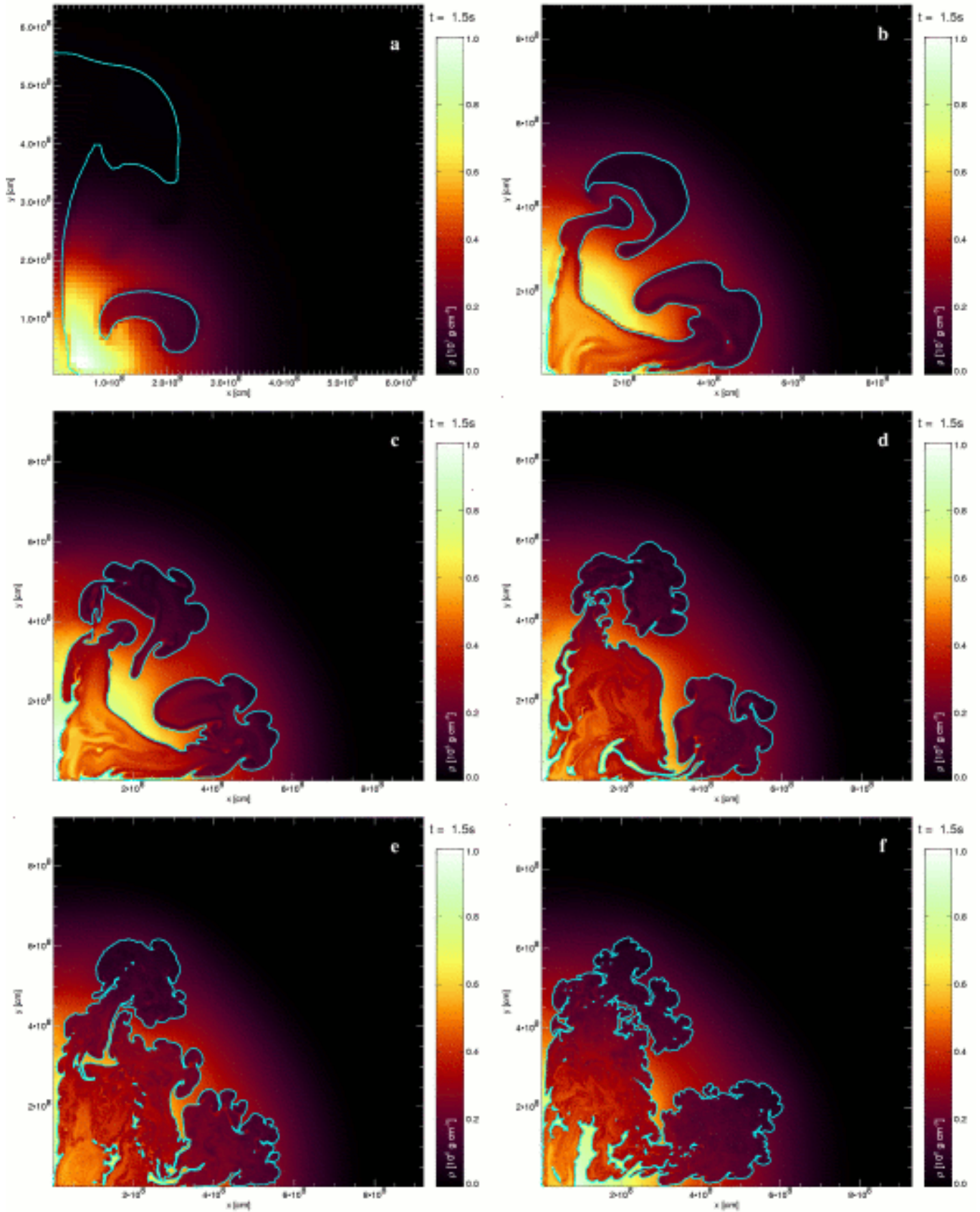}
}
\caption{Two-dimensional SN Ia simulations with different domain
  discretizations: \textbf{(a)} $[64]^2$, \textbf{(b)} $[128]^2$,
  \textbf{(c)} $[256]^2$, \textbf{(d)} $[512]^2$, \textbf{(e)}
  $[1024]^2$, and \textbf{(f)} $[2048]^2$ cells. \label{sn2d_fig}}
\end{figure*}

Numerical convergence is a basis to judge on the credibility of
simulations. 
In the scenario under consideration, the flame front is unstable and
affected by turbulence on scales that reach far below the
resolution that is achievable in multi-dimensional simulations. From
this point of view, numerical convergence in the details of the flame
morphology cannot be reached.
However, in our models convergence in global quantities (like the
total energy generation or the production of burning products) is
expected to result from the interplay of large-scale flame front features
with the
sub-grid scale model. Ideally, a lack of resolution of
large-scale features in the flame front representation should be
compensated by an increased turbulent flame propagation velocity
determined by the sub-grid scale approach. Of course, a certain
threshold of resolution will need to be exceeded to reach this regime
in our numerical implementation.

Our final goal is to simulate SNe Ia
in three spatial dimensions. However, currently it is computationally too
expensive to perform three-dimensional convergence tests. We
therefore take the approach of \citet{reinecke2002c} and perform the
resolution study in two dimensions. From this we conclude on the
behavior in three dimensions. For our convergence tests we conducted a
series of two-dimensional simulations with resolutions of $[64]^2$,
$[128]^2$, $[256]^2$, $[512]^2$, $[1024]^2$, and $[2048]^2$ grid cells. The
flame fronts and densities after $1.5 \, \mathrm{s}$ (when burning has
ceased) are visualized in
Fig.~\ref{sn2d_fig}. The model with the lowest resolution
(Fig.~\ref{sn2d_fig}a) shows a
peculiar behavior. 
In the better resolved simulations the large-scale features resemble
each other to a certain degree. But also here the
details differ. The buoyancy instability
of the flame front gives rise to a nonlinear evolution of the
large-scale flame features. The particular realization of this is
sensitive to small perturbations and thus the different numerical
noise in the models may lead to variations in the
large-scale features. Thus, even in numerically converged models a
small scatter of the global characteristics is possible.

For the static grid simulation
\citet{reinecke2002c} found numerical convergence in the energy
production for grid spacings of the inner part of the domain below
$\Delta x \approx 10^6 \, \mathrm{cm}$. However, for the late stages,
when the flame front enters the 
nonuniform part of the domain, the energy production failed to
converge in that test.

\begin{figure}[t]
\centerline{
\includegraphics[width = \linewidth]
  {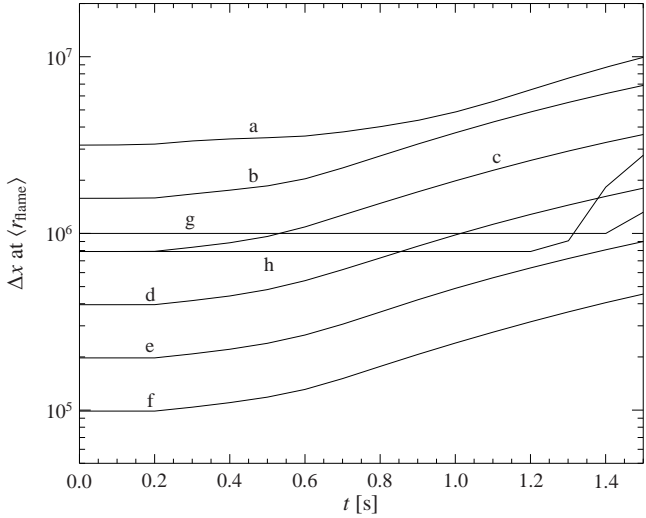}}
\caption{Resolution of the flame front in co-expanding simulations
  with (a) $[64]^2$ cells, (b) $[128]^2$ cells, (c)
  $[256]^2$ cells, (d) $[512]^2$ cells, (e) $[1024]^2$ cells, (f) $[2048]^2$ cells, and in
  static grid simulations with (g) $[256]^2$ cells and an inner grid
  spacing of $10^6 \, \mathrm{cm}$,
  (h) $[256]^2$ cells and an inner grid spacing of $7.9 \times 10^5 \,
  \mathrm{cm}$. \label{fl_res_fig}} 
\end{figure}

This points to an important issue in our simulations.
Neither in the co-expanding nor in the static grid approach
the resolution of the flame front is constant.
For the latter the
flame front resolution is constant only in the early stages of the explosion but
decreases rapidly once the flame enters the non-uniform part of the
computational grid. In the co-expanding grid approach, the resolution
of the flame front decreases steadily. This is
demonstrated in Fig.~\ref{fl_res_fig}, where the resolution of the
flame front is
plotted against time. In this plot, graph g
corresponds to a static grid simulation that has reached numerical
convergence according to \cite{reinecke2002c}. It is obvious that only
co-expanding grid simulations with $[1024]^2$ or more grid cells resolve
the flame front better at all times. It needs to be emphasized at this
point, that for the co-expanding grid model the situation is further
complicated by the fact that the grid resolution is adapted to the
expansion of the star. Thus, more vigorous explosions will decrease the
numerical resolution faster.

\begin{figure}[t]
\centerline{
\includegraphics[width = \linewidth]
  {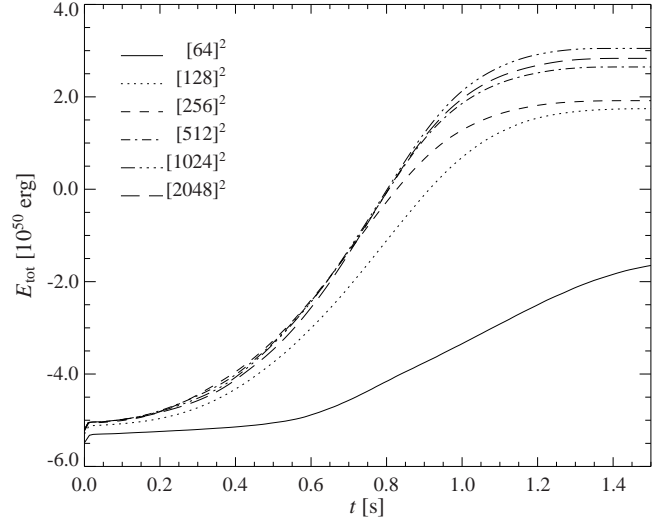}
}
\caption{Total energy in simulations with co-expanding grid for
  different numbers of computational grid cells. \label{etot_all_fig}}
\end{figure}

The total energies produced in our simulations with different domain
discretizations are plotted in Fig.~\ref{etot_all_fig}. The runs with
$[64]^2$ and $[128^2]$ cells show a peculiar
behavior. The simulations with a resolution of $[512]^2$ cells and better
seem to be numerically converged. It is interesting to note that the
model with $[2048]^2$ cells releases slightly less energy than the
$[1024]^2$ cells model. As discussed above, this can be attributed to
the slightly different large-scale realization of the flame front
which is obvious from a comparison of Figs.~\ref{sn2d_fig}e and
\ref{sn2d_fig}f.
The model with a resolution of $[256]^2$ cells is converged in the
early stages but the energy production falls behind the higher
resolved runs for $t \lesssim 0.7 \, \mathrm{s}$. This is similar 
to what was found by \citet{reinecke2002c} and can be
attributed to the decreasing resolution of the flame front. 

\begin{figure}[t]
\centerline{
\includegraphics[width = \linewidth]
  {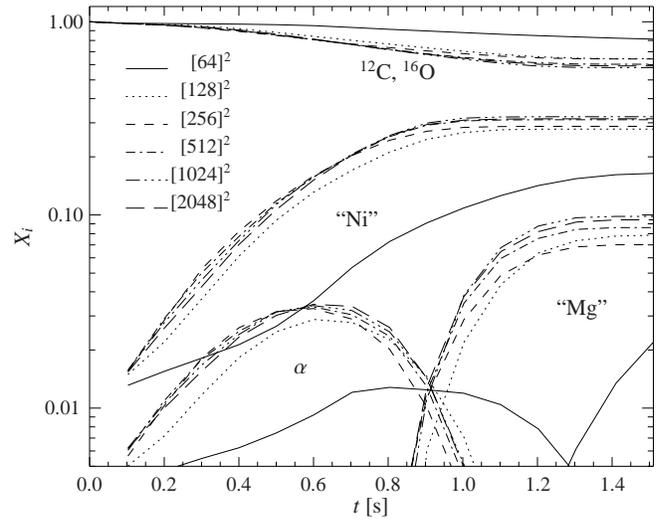}
}
\caption{Evolution of the chemical composition in simulations with co-expanding grid for
  different numbers of computational grid cells. \label{evalmass_fig}}
\end{figure}

\begin{figure*}[t!]
\centerline{
\includegraphics[width = \linewidth]
  {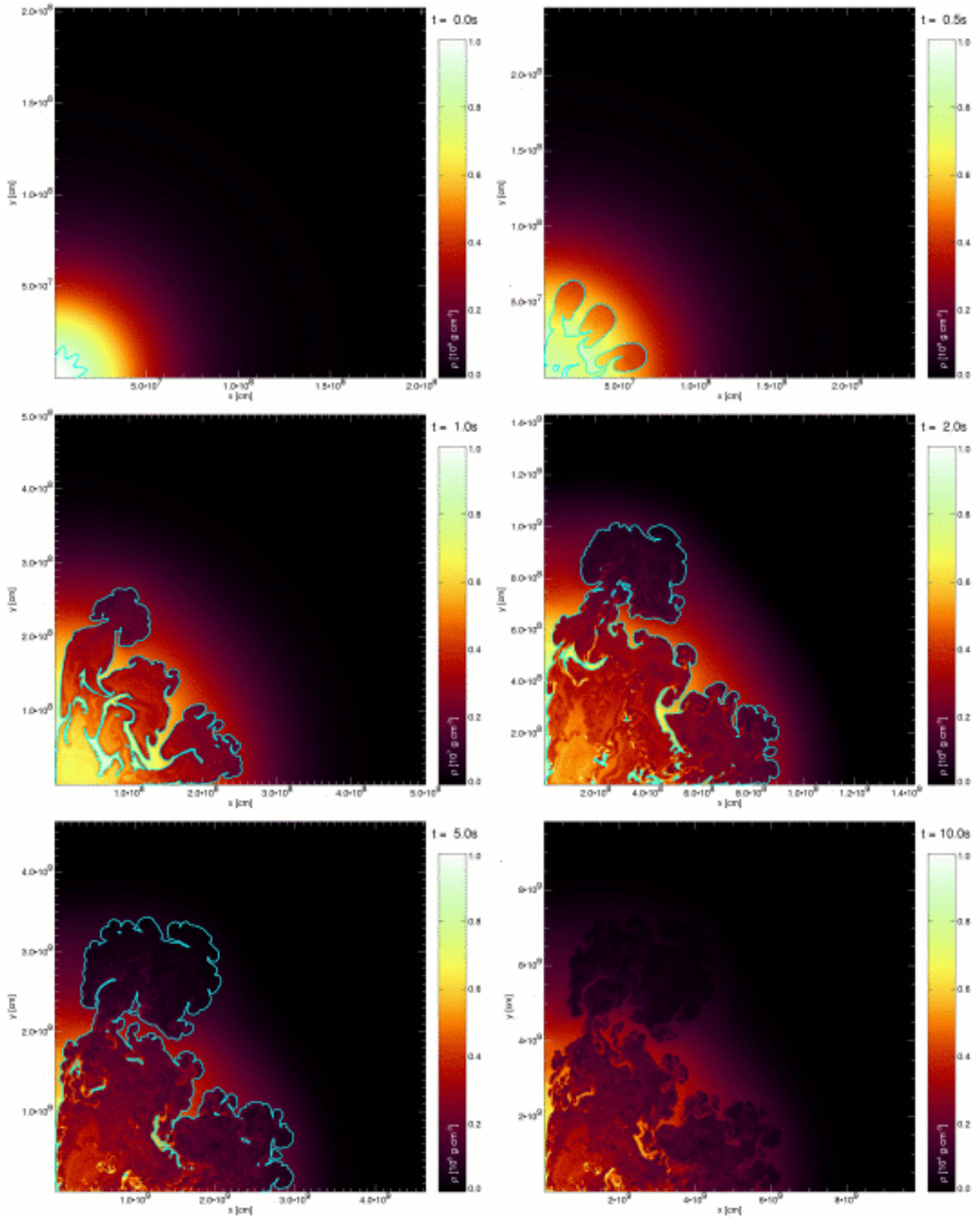}
}
\caption{Two-dimensional SN Ia simulation with $[1024]^2$
  cells. \label{evo_fig}}
\end{figure*}

The evolution of the chemical composition in our simulations is
determined by the propagation velocity of the flame front and the fuel
density at the flame front and thus closely related to the expansion
of the WD, which in turn depends on the flame front propagation. The
temporal evolution of the composition is plotted in
Fig.~\ref{evalmass_fig}. It is obvious that in the models with $[64]^2$
and $[128]^2$ cells the chemical evolution deviates considerably from
the better resolved runs. Here an insufficiently resolved flame front burns
less material but also leads to a slower expansion of the star (cf.\
Fig.~\ref{sn2d_fig}). For the other models the production of iron
group elements can be regarded as converged but the $[256]^2$ cells
simulation shows less intermediate mass elements, which are produced
in later stages of the explosion. 

This makes the simulation with $[256]^2$ cells the most important to
judge on numerical convergence. Both the evolution of the total energy
and the chemical composition in that model suggest that the resolution
of the flame front becomes too coarse to be numerically resolved at $t
\approx 0.7 \, \mathrm{s}$. According to Fig.~\ref{fl_res_fig} this
corresponds to a resolution of  $\sim$$15 \, \mathrm{km}$.

Our study implies that for
two-dimensional simulations this is given for models with more than
$[512]^2$ cells starting from a resolution of better than $3.95 \, \mathrm{km}$. A
model with $[256]^2$ (with ($\Delta x)_\mathrm{ini} = 7.9 \,
\mathrm{km}$) cells will give the correct production of iron
group elements but slightly underproduce intermediate mass
nuclei. Since three-dimensional simulations explode stronger than
their two-dimensional counterparts (cf. \citealt{reinecke2002b}), a
somewhat better initial resolution is desirable here. 
It needs to be emphasized that the resolution is not only crucial for 
numerical convergence, but also for the implementation of the physical
model. \cite{reinecke2002d} find the most energetic
explosions for multi-spot ignition scenarios. To accommodate an adequate
number of initial flame spots, one needs, however a rather fine
computational grid.

\subsection{Reaching homologous expansion}

\begin{figure}[t]
\centerline{
\includegraphics[width = \linewidth]
  {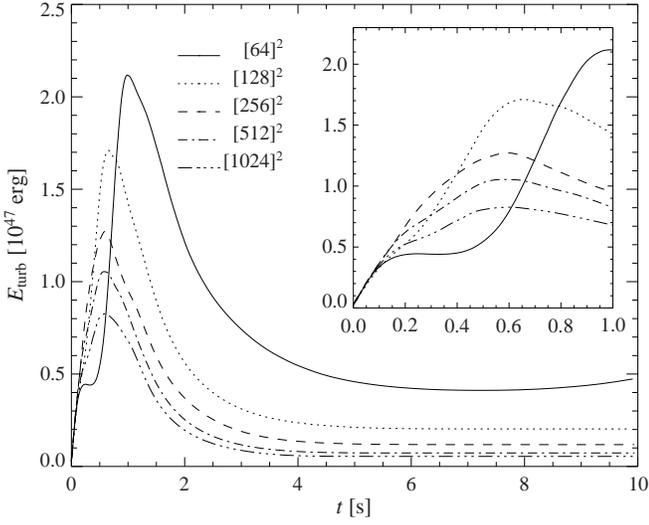}
}
\caption{Evolution of the turbulent energy in simulations with
  co-expanding grid for 
  different numbers of computational grid cells. \label{eturb_all_fig}}
\end{figure}

Figure~\ref{evo_fig} depicts the temporal evolution of our SN Ia
simulation with $[1024]^2$ cells up to $t=10.0\, \mathrm{s}$. Note
that the color-coded density range varies in the snapshots. The solid
contour represents the zero level set of the $G$-function. This is
associated with the flame front at early times, but after $t\sim 2\,
\mathrm{s}$ the density of the WD has become so low due to expansion
that burning ceases in our description. It is, however, interesting to
note that the $G=0$ level set is still a good indicator of the
interface between unburnt and burnt material at later times (see
snapshot at $t=5.0\, \mathrm{s}$ and compare to the snapshot at
$t=10.0 \, \mathrm{s}$ where the $G=0$ contour was omitted). 
The reason for this somewhat surprising behavior is that the turbulent
energy in our simulation decreases drastically after about $2 \,
\mathrm{s}$ This is shown in Fig.~\ref{eturb_all_fig} where the
total sub-grid scale turbulent energy is plotted against time.
One has to keep in mind,
however, that $q_\mathrm{sgs}$ increases with the size of the grid
cells, which even moderates the decrease of the turbulent energy plotted in
Fig.~\ref{eturb_all_fig}.
Hence later
than $t \approx 2 \, \mathrm{s}$ 
$q_\mathrm{sgs}$ becomes so small that the fluid velocity $\mbf{v}$
dominates in Eq.~(\ref{flame_vel_eq}) and consequently
the $G = 0$ isosurface (isoline) is
advected passively with the resolved flow.

Figure~\ref{eturb_all_fig} again demonstrates numerical convergence for
our models with more than $[256]^2$ cells. Here the criterion is not
an exact match of the value of the turbulent sub-grid scale energy but
its temporal evolution.
This is similar for the highly resolved simulations while the less
resolved runs diverge at early
times. 

The
morphology of the density structures does not significantly change
after $t \sim 5 \, \mathrm{s}$. This indicates that our simulation
approaches homologous expansion.
How consistent our simulation at late times really is with the
assumption of homologous expansion has to be decided on the basis of
the velocity profile and the amplitude of the gradients of pressure
and the gravitational potential.

\begin{figure}[t]
\centerline{
\includegraphics[width = \linewidth]
  {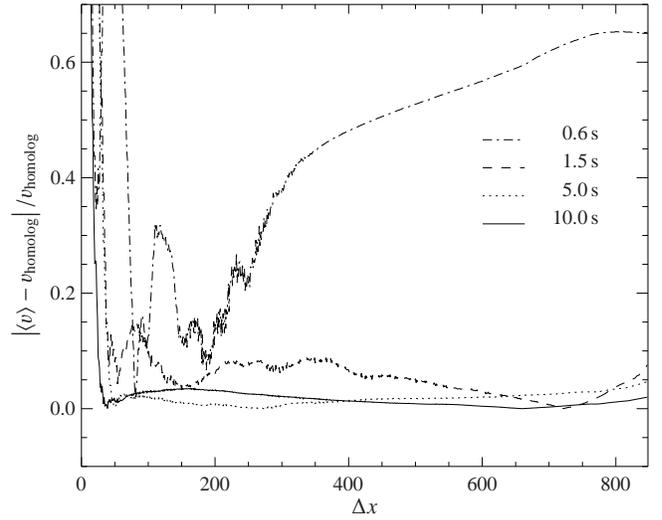}
}
\caption{Deviation of the angular averaged velocity from homologous
  expansion at different times in the two-dimensional simulation with
  $[1024]^2$ cells. \label{homolog_v_fig}}
\end{figure}

This will be analyzed for our  simulation
with $[1024]^2$ cells.
Fig.~\ref{homolog_v_fig} shows the deviation of the angular averaged
velocities from relation (\ref{homolog_eq}) at different times. The
proportionality factor of that relation was obtained from a fit to the velocity profile
at $t=10 \, \mathrm{s}$.
The velocity deviations are plotted against the distance from the WD's
center. This distance is expressed in
widths of computational cells $\Delta x$ to allow comparison; because
of the uniform grid the physical distance scales directly with $\Delta
x$. We note that the deviations from homologous expansion decrease
drastically. Except from the region in the very center the deviation
at $t=10 \, \mathrm{s}$ stays below 5\%. The static grid simulations
ended at $t = 1.5 \, \mathrm{s}$. Here deviations still reach
10\%. The maximal values and the 
mean values of the pressure gradients are given in
Table~\ref{pg_tab}. These quantities also decrease drastically with
time. From $t=0.6 \, \mathrm{s}$ (where the energy generation rate
peaks) to $t=1.5 \, \mathrm{s}$ they differ less than a magnitude
while from  $t=1.5 \, \mathrm{s}$ to $t = 10.0\, \mathrm{s}$ they drop
two magnitudes. The evolution of the gradients of the gravitational
potential (see Table~\ref{pg_tab}) shows the same trend.

\begin{figure}[t]
\centerline{
\includegraphics[width = \linewidth]
  {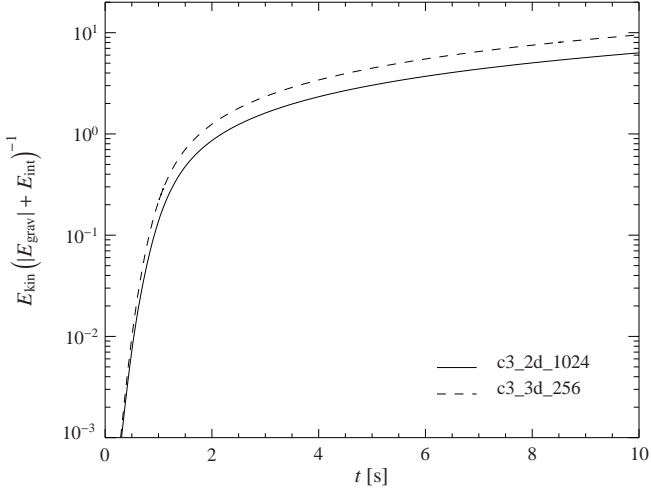}
}
\caption{Temporal evolution of the ratio of the kinetic energy to the
  sum of internal and gravitational energy for the two-dimensional
  model with $[1024]^2$ cells (solid curve) and the three-dimensional
  model with $[256]^2$ cells (dashed curve). \label{homolog_e_fig}}
\end{figure}

We conclude that our simulations at $t \approx 10
\, \mathrm{s}$ provide a fair approximation to homologous expansion.
This is supported by Fig.~\ref{homolog_e_fig} (solid curve) showing
that at $t \approx 10
\, \mathrm{s}$ the kinetic energy of our model is almost one order of
magnitude larger than the sum of the internal and gravitational energies.

\begin{table}
\centering
\caption{Pressure gradients and gradients of the gravitational
  potential in the two-dimensional $[1024]^2$ cells simulation at
  different times.
\label{pg_tab}}
\setlength{\extrarowheight}{2pt}
\begin{tabular}{rp{0.2 \linewidth}p{0.2 \linewidth}p{0.2 \linewidth}}
\hline\hline
$t$ [s] & $\max \left| \mbf{\nabla} p \right|$
$[\mathrm{dyn} \, \mathrm{cm}^{-3}]$ &$\langle \left| \mbf{\nabla} p
\right| \rangle$ 
$[\mathrm{dyn} \, \mathrm{cm}^{-3}]$ & $\max \left| \mbf{\nabla} \Phi \right|$
$[\mathrm{cm} \, \mathrm{s}^{-2}]$
\\
\hline
0.6  & $9.44 \times 10^{7}$ & $1.17 \times 10^{7}$ & $8.17 \times 10^{9}$\\
1.5  & $6.85 \times 10^{7}$ & $2.95 \times 10^{6}$ & $4.05 \times 10^{8}$\\
5.0  & $3.82 \times 10^{5}$ & $8.74 \times 10^{4}$ & $1.74 \times 10^{7}$\\
10.0 & $1.27 \times 10^{5}$ & $3.40 \times 10^{4}$ & $4.09 \times 10^{6}$\\
\hline
\end{tabular}
\end{table}

Due to the co-expanding grid the computational costs to reach
homologous expansion are moderate. The expanding grid spacing leads to
an increase in the time steps satisfying the CFL criterion. For the
Godunov-scheme used to solve the 
equations of hydrodynamics in our implementation this criterion states
for the time step $\Delta t$
$$
\Delta t \le \Delta x \left( \mathrm{max}_i \left\{ |\mbf{v}_i| +
  c_{\mathrm{s},i} \right\}\right)^{-1},
$$
where the index $i$ runs over all computational cells and $\mbf{v}$
and $c_\mathrm{s}$ denote the fluid velocity and the sound speed,
respectively. The CFL criterion is necessary but not sufficient to
ensure stability of the scheme. To guarantee stability we reduce the
time step by a factor of $0.8$ in our implementation. 

\begin{figure}[t]
\centerline{
\includegraphics[width = \linewidth]
  {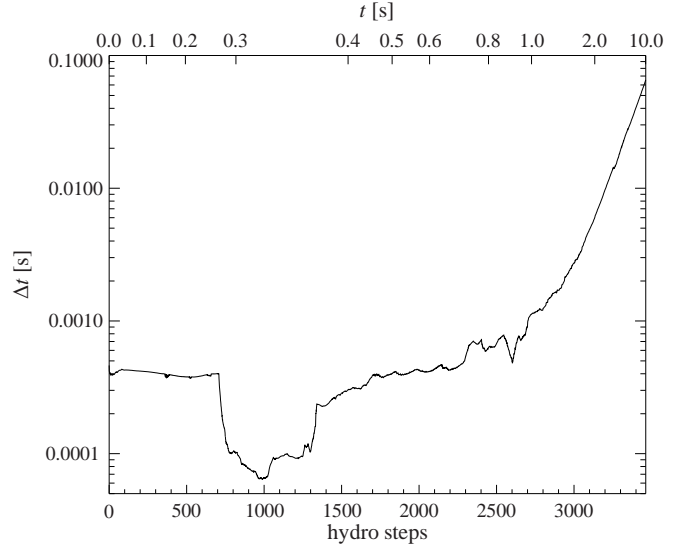}
}
\caption{Time steps in the simulation with $[256]^2$ cells. \label{cfl_fig}}
\end{figure}

The evolution of
the time steps taken in our simulation with $[256]^2$ cells is plotted
in Fig.~\ref{cfl_fig}. Two points should be noted here. First, the
time step is very low between $\sim$$0.30\, \mathrm{s}$ and
$\sim$$0.38\, \mathrm{s}$. The reason for this is that the shock wave
generated by the initial flame setup reaches the WD's surface and
propagates into the surrounding pseudo-vacuum. Here high velocities
are generated so that computational cells outside the star dominate
in the CFL criterion. However, once the first shock wave has crossed
the WD, it starts to expand and the computational grid co-expands
accordingly. With the reflecting outer boundaries this leads to a
damping of the motions in the 
cells filled by the pseudo-vacuum and the time step recovers to
larger values. The second point to note here is that the increase of
the time step due to the expansion makes it possible to reach
from $2 \, \mathrm{s}$ to $10 \, \mathrm{s}$ with only $\sim$$10\%$
more hydro steps than calculating up to $2 \, \mathrm{s}$.

\section{A three-dimensional simulation}

\begin{figure*}[p]
\centerline{
\includegraphics[height = 0.95 \textheight]
  {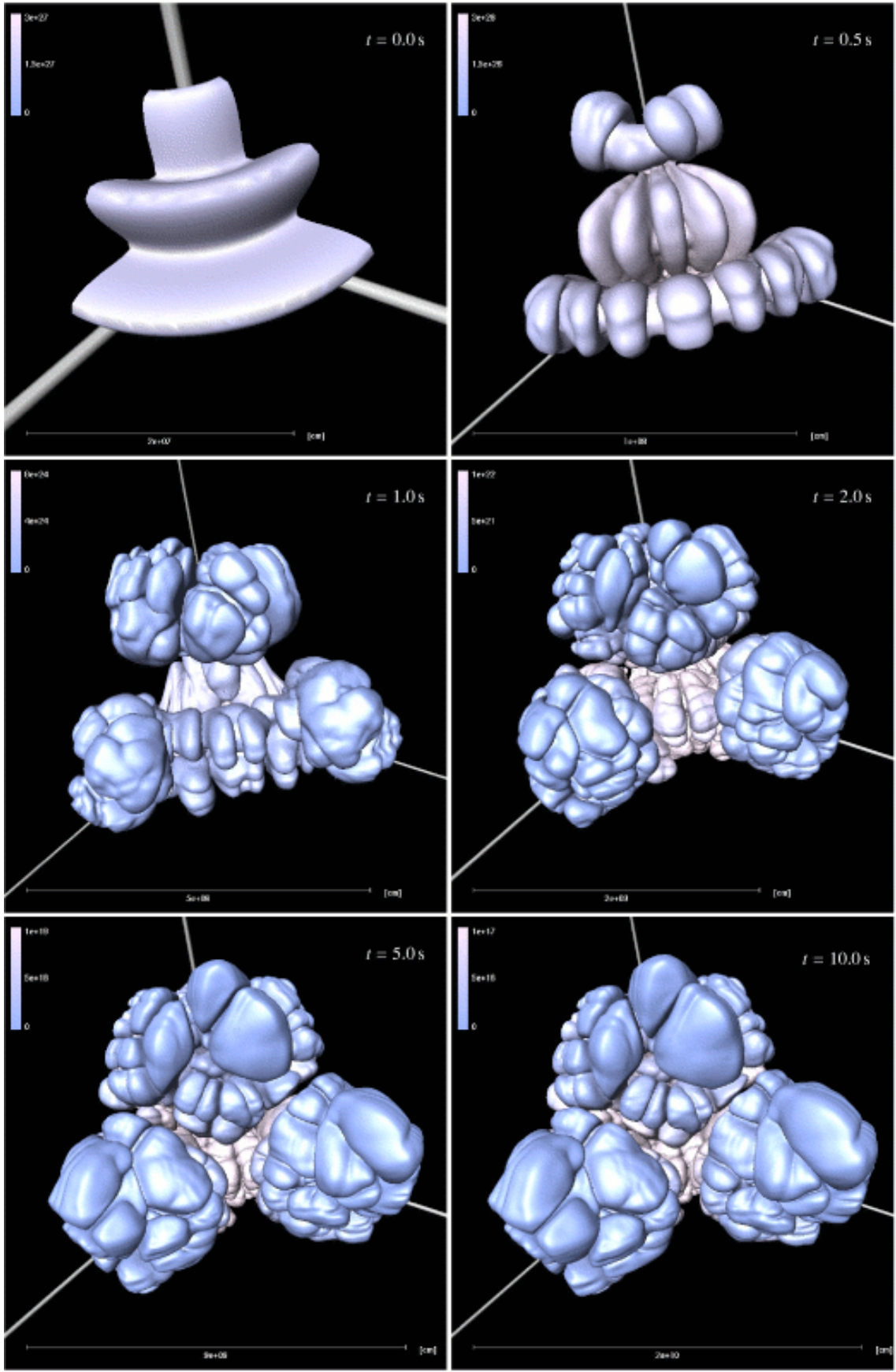}
}
\caption{Three-dimensional SN Ia simulation. The isosurface
  corresponds to $G = 0$ and the color-coding indicates the value of
  the sub-grid scale energy providing a measure of the propagation
  velocity of the front. The scale gives an impression of the
  expansion and corresponds to the plane through
  the origin of the coordinate system.\label{sn3d_fig}}
\end{figure*}

With the scheme described in the previous sections we performed a
three-dimensional test simulation. The computational domain was
discretized into $[256]^3$ cells starting with a grid spacing of $7.9
\times 10^5 \, \mathrm{cm}$
and the flame setup corresponded to the three-dimensional version of
the \emph{c3}-model by \citet{reinecke2002d, reinecke2002c}. Snapshots from the
simulations are shown in Fig.~\ref{sn3d_fig}. The rendered isosurface
corresponds to the zero level set of $G$. As discussed above it
represents the flame front at early stages and still provides an
approximate separation between fuel and ashes after
$t\sim2\,\mathrm{s}$.

The early evolution of the flame front is very similar to that
found by \citet{reinecke2002d, reinecke2002c}. The initial axial
symmetry is broken in the same way. The late stages show more
structure due to the better resolution of the co-expanding grid in the
outer parts of the domain.
Our study in the previous section indicated that the simulation is close
to homologous expansion already after $2\,\mathrm{s}$. In the
three-dimensional simulation, the morphology of the $G = 0$
isosurface still changes. The outer bubbles grow hiding the inner
structures. This may have impact on the light curves and the spectra
derived from such models. From $t=5\, \mathrm{s}$ to
$t=10\, \mathrm{s}$ the visible changes are only marginal.

\begin{figure}[t]
\centerline{
\includegraphics[width = \linewidth]
  {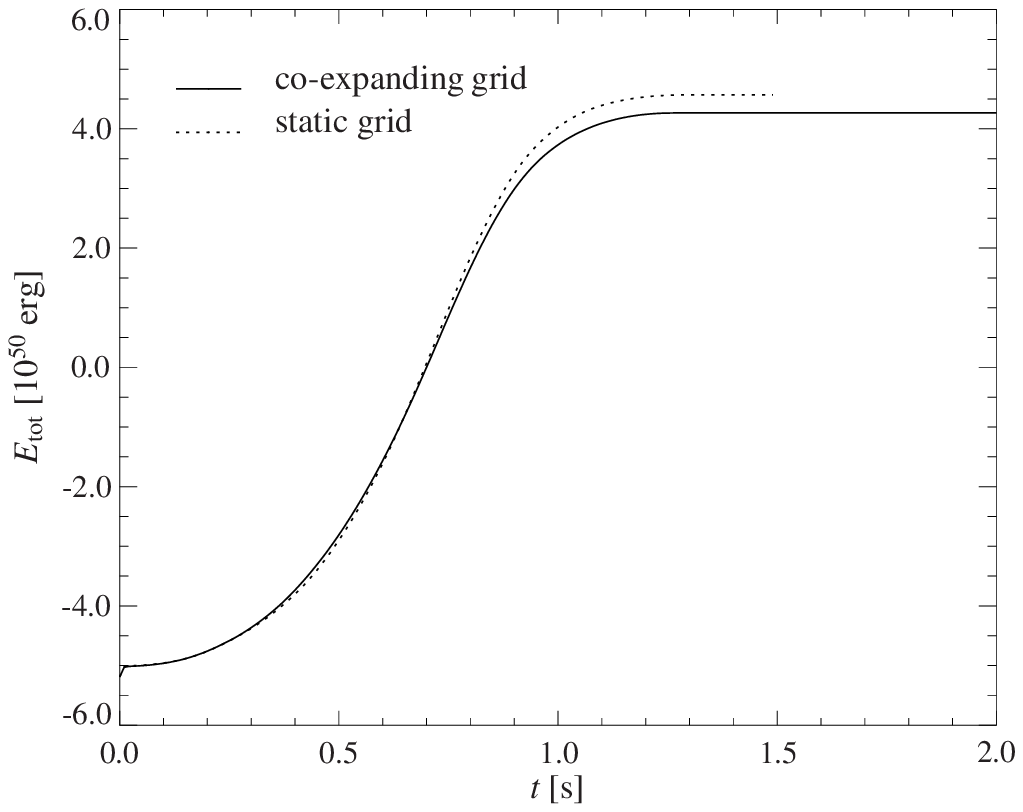}
}
\caption{Comparison of the total energies of our three-dimensional
  simulation with co-expanding grid and the \emph{c3\_3d\_256}
  model (static
  grid) by \citet{reinecke2002d, reinecke2002c}. \label{etot_compare_3d_fig}}
\end{figure}

The energetics of the model is compared to the  \emph{c3\_3d\_256}
model\footnote{The data was kindly provided by M.~Reinecke. In this
model the grid spacing at the inner part of the computational domain
was $\Delta x = 10^6 \, \mathrm{cm}$.} by \citet{reinecke2002d, reinecke2002c} in
Fig.~\ref{etot_compare_3d_fig}. As in 
the two-dimensional simulations the differences between the
static grid and the co-expanding grid simulations can be attributed
to the varying resolution of the flame front in the latter.

\begin{figure}[t]
\centerline{
\includegraphics[width = \linewidth]
  {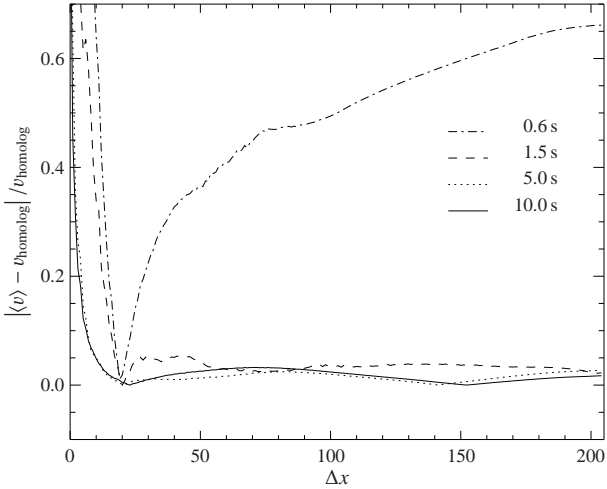}
}
\caption{Deviation of the angular averaged velocity from homologous
  expansion at different times in the three-dimensional
  simulation. \label{homolog3d_v_fig}} 
\end{figure}

\begin{figure}[t]
\centerline{
\includegraphics[width = \linewidth]
  {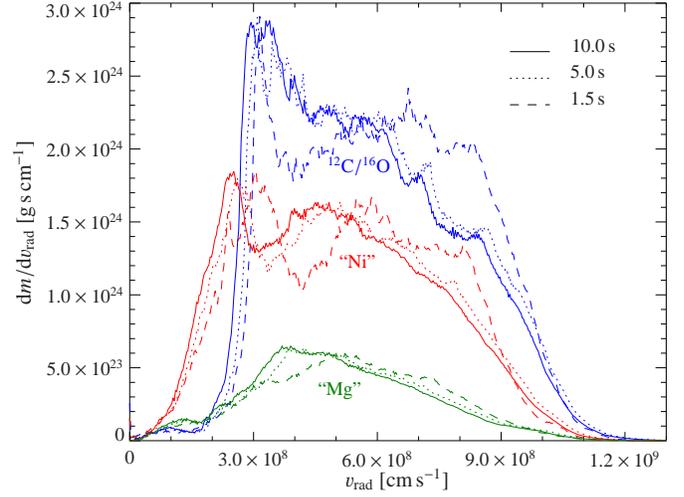}
}
\caption{Distribution of the species in radial velocity
  space. ``Ni'' and ``Mg''  denote the iron group elements and the
  intermediate mass elements, respectively. \label{velprofile_fig}} 
\end{figure}

Fig.~\ref{homolog3d_v_fig} shows the deviation from a self-similar
velocity profile in our three-dimensional simulation. We note that it
approaches a relation consistent with Eq.~(\ref{homolog_eq}) faster
than the two-dimensional simulations. This can at least partly be
attributed to the more energetic explosion resulting in three
dimensions, giving rise to higher expansion velocities. 
Figure~\ref{homolog_e_fig} corroborates this interpretation
showing an accelerated conversion of the
energy forms to kinetic energy in the three-dimensional model. The
gradients of the pressure and the gravitational potential are given in
Table~\ref{pg3d_tab}. The maximum value of $| \mbf{\nabla} \Phi |$
develops similar to the two-dimensional case while the
mean of the pressure gradient amounts to roughly half the value
obtained in the two-dimensional simulation. As in the two-dimensional
model, the pressure gradient drops significantly between $t = 1.5 \,
\mathrm{s}$ and  $t = 5.0 \, \mathrm{s}$.

\begin{table}
\centering
\caption{Pressure gradients and gradients of the gravitational
  potential in the three-dimensional simulation at
  different times.
\label{pg3d_tab}}
\setlength{\extrarowheight}{2pt}
\begin{tabular}{rp{0.2 \linewidth}p{0.2 \linewidth}p{0.2 \linewidth}}
\hline\hline
$t$ [s] & $\max \left| \mbf{\nabla} p \right|$
$[\mathrm{dyn} \, \mathrm{cm}^{-3}]$ &$\langle \left| \mbf{\nabla} p
\right| \rangle$ 
$[\mathrm{dyn} \, \mathrm{cm}^{-3}]$ & $\max \left| \mbf{\nabla} \Phi \right|$
$[\mathrm{cm} \, \mathrm{s}^{-2}]$
\\
\hline
0.6  & $5.49 \times 10^{7}$ & $4.02 \times 10^{6}$ & $6.93 \times 10^{9}$\\
1.5  & $5.57 \times 10^{6}$ & $1.50 \times 10^{6}$ & $3.16 \times 10^{8}$\\
5.0  & $1.45 \times 10^{5}$ & $4.74 \times 10^{4}$ & $1.49 \times 10^{7}$\\
10.0 & $5.70 \times 10^{4}$ & $1.90 \times 10^{4}$ & $3.59 \times 10^{6}$\\
\hline
\end{tabular}
\end{table}

The distributions of chemical species in radial velocity space for
different times are shown in Fig.~\ref{velprofile_fig}. The changes
between $t=5.0\, \mathrm{s}$ and $t=10.0\, \mathrm{s}$  are
small. The profile is somewhat shifted toward smaller velocities which
corresponds to a deceleration of the ejecta. This is not surprising since
the gravitational potential is small, but not vanishing. In contrast,
between $t=1.5\, \mathrm{s}$ and $t=10.0\, \mathrm{s}$ we observe
substantial changes in the profiles. All three species tend to pile up
at lower velocities diluting the high-velocity components.

We conclude that our scheme is applicable to three-dimensional
simulations. The expansion of the computational grid according to the
tracked $1.4 \, M_\odot$ mass shell did not cause any
difficulties. Homologous expansion in three-dimensional simulations
will be reached somewhat quicker than in two-dimensional models.

\section{Conclusions}

We analyzed the approach of multi-dimensional SN Ia models to
homologous expansion. To this end a novel scheme with co-expanding
computational grid was implemented and tested. Taking into account the
variable resolution resulting from this scheme, a comparison with
previous simulations on a static computational grid showed good
agreement in the flame evolutions. 

A convergence study validated our co-expanding grid approach. Here,
the variability of the numerical resolution leads to the conclusion
that the grid spacing has to stay below $\sim$$15 \, \mathrm{km}$ for
numerically converged stages of the explosion. For two-dimensional
simulations one has to start with a domain discretization finer than
$[512]^2$ cells to resolve the whole explosion process, but a
$[256]^2$ cell models will already provide the correct production of
iron group elements. In case of three-dimensional explosion models a
somewhat finer initial resolution is required because the explosions
here are more vigorous resulting in a faster expansion.

Applying our co-expanding grid scheme, it was possible to follow the
explosion significantly longer than in any previous multi-dimensional SN
Ia simulation with only marginal additional computational
expense. From the temporal evolution of the velocity profiles, the energy
conversion and the gradients of pressure and gravitational potential
we conclude that homologous expansion is reached to a high accuracy at
$t \sim 10 \, \mathrm{s}$. Although the velocity profile is rather
quickly in agreement with Eq.~(\ref{homolog_eq}), the pressure
gradient drops significantly at later times. Therefore the flame
morphology still changes until $t \sim 5 \, \mathrm{s}$. Outer bubbles
grow and hide inner parts of the ejecta. Between $t = 1.5 \,
\mathrm{s}$ and $t=10 \, \mathrm{s}$ we observe a significant
redistribution of chemical species in velocity space.
Thus, for the derivation of synthetic light curves and spectra from
SN Ia models it is advisable to follow the evolution of the explosion
up to $t \sim 10 \, \mathrm{s}$.

The novel numerical scheme presented here offers a variety of
improvements of multi-dimensional SN Ia explosion models. A uniform
cartesian grid alleviates some numerical problems (see
Sect.~\ref{intro_sect}) and moreover enables the implementation of
more sophisticated sub-grid scale models (Schmidt et al., in
preparation). Full star SN Ia explosion models, where the artifical
symmetry to the spatial octants is abolished, also gain accuracy
from such a computational grid \citep{roepke2004d}. It is, however,
also possible to combine the co-expanding grid approach with the
highly non-uniform grid used in previous static-grid simulations. In
this way a high resolution of the flame front (especially in the
early stages) will be possible with low computational expenses and
initial flame conditions can be tested easily. We will address this
question in a subsequent study.

An open issue in the current implementation is the
modeling of flame propagation in late stages of the explosion. Here
the finite width of the flame is not negligible anymore since
turbulent eddies start to penetrate the preheat zone of the flame. The
so-called flamelet assumption of turbulent combustion breaks down and
the flame enters the thin reaction zones regime \citep[for a discussion of
the turbuelnt burning regimes see e.g.][and for
application to SN Ia explosion see \citealp{niemeyer1997d}]{peters1999a}. It is,
however, unclear whether burning in this regime contributes
significantly to the explosion results. This
issue will be addressed in a forthcoming study.


\begin{acknowledgements}
The author gratefully announces helpful discussions with
W.~Hillebrandt, M.~Reinecke, 
K.~Kifonidis, and W.~Schmidt concerning the numerical implementation
and astrophysical modeling. L.~Scheck
helped with the routines to visualize the 2D color snapshots of our
simulations. This work was supported in part by the European Research
Training Network ``The Physics of Type Ia Supernova Explosions'' under
contract HPRN-CT-2002-00303.
\end{acknowledgements}

\end{document}